\pdfoutput=1
\documentclass[]{raa}            

\usepackage{graphicx,times}             
\usepackage{natbib}
\usepackage{amssymb,amsmath}
\bibpunct{(}{)}{;}{a}{}{,}

\usepackage[pagebackref=true]{hyperref}

\begin{document}

\title{\ Investigating the coincidence problem with Interacting Dark Energy Models}

\volnopage{Vol.0 (20xx) No.0, 000--000}      
   \setcounter{page}{1}          

 \author{Kshitiz Singh and Savita Gahlaut
   \inst{}
   }

   \institute{Deen Dayal Upadhyaya College, University of Delhi, Sector-3, Dwarka, New Delhi 110078, India \\
 E-Mail: savitagahlaut@ddu.du.ac.in  } 
      
\vs\no
   {\small Received 20xx month day; accepted 20xx month day}   



\abstract{
Interacting dark energy (IDE) models within the framework of general relativity (GR) provide a possible extension of the standard $\Lambda$CDM model and may alleviate the coincidence problem.
In this paper, we investigate three phenomenological interacting dark energy models using a combination of cosmic chronometer $H(z)$ measurements and two standard cosmic rulers: baryon acoustic oscillations (BAO) and quasars (QSO). The observational constraints indicate a mild preference for energy transfer from dark energy to dark matter, a scenario that may help alleviate the cosmic coincidence problem. However, the standard $\Lambda$CDM model, corresponding to the absence of interaction, remains fully consistent with the observational data, as the no-interaction case lies within the $1\sigma$ confidence interval of the inferred interaction parameter for all three IDE models. Model comparison, using  Akaike Information Criterion (AIC)  and the Bayesian Information Criterion (BIC), further indicates that the current data do not provide  statistically significant evidence favouring  any  IDE model over the standard $\Lambda$CDM scenario. These results suggest that, while a weak dark-sector interaction is permitted by the data, present observations do not provide compelling evidence for departures from the standard $\Lambda$CDM  model.
\keywords{Cosmology: Observations --- Cosmology: Dark Energy --- Cosmology: Physical Data and Processes --- Cosmology: Distance Scales}
}

 \authorrunning{K. Singh $\&$ S. Gahlaut }            
   \titlerunning{IDE Models }  

   \maketitle

\section{Introduction}
\label{sect:intro}

Observations of Type Ia supernovae (SNe Ia) by \citet{Reiss} and \citet{Perl} provided the first compelling evidence that the Universe at present is undergoing accelerated expansion. Subsequent and independent probes, for example:  measurements of the cosmic microwave background radiation (CMBR), baryon acoustic oscillations (BAO), and cosmic chronometers through $H(z)$ determinations, have consistently supported this conclusion. The current accelerated expansion of the Universe is generally attributed to the effects of  dark energy , a mysterious component whose origin and nature remain unknown. 
The $\Lambda$CDM model, which  best accommodates the broad range of current observational data, has emerged as the concordance model and is widely regarded as the standard model of modern cosmology \citep{Peebles}. In the $\Lambda$CDM model, the cosmological constant, $\Lambda$, represents the dark energy component, which is assumed to be spatially homogeneous and constant in time. Dark energy, modelled by $\Lambda$, constitutes approximately $70\%$ of the current energy density of the Universe, while cold dark matter (CDM) accounts for about $25\%$, making these two components the dominant constituents of the current cosmic energy budget.
 
Over the past few decades, both the precision and volume of cosmological observations have increased significantly. While this progress has strengthened the empirical foundations of modern cosmology, it has also revealed a number of persistent anomalies and internal inconsistencies across different datasets. Prominent among these are the Hubble tension; the disagreement between independent determinations of the Hubble constant based on observations of the early Universe and those derived from late-time cosmological measurements, the $S_{8}$ tension associated with the estimates of the clustering amplitude parameter of matter fluctuations, and certain large-scale anomalies or unexpected features in the CMBR anisotropy measurements \citep{Verde,Joseph,Schwarz} .

In addition to these observational tensions, the standard $\Lambda$CDM model faces
several conceptual and theoretical challenges. 
The dark energy, in $\Lambda$CDM model, is in the form of vacuum energy, characterized by  
the cosmological constant $\Lambda$. However, the value of dark energy density estimated from the observations differ from theoretical estimates based on quantum field theory by about 120 orders of magnitude, giving rise to the well-known cosmological constant problem \citep{Weinberg}. In addition,  observations indicate  that the present-day dark energy density parameter, $\Omega_{\Lambda}$, and the cold dark matter energy density parameter, $\Omega_{m}$,  are of comparable magnitude, despite the fact that they evolve very differently over cosmic time. This apparent fine-tuning constitutes the coincidence problem \citep{Zlatev}.
These conceptual challenges, together with the observational tensions, continue to motivate the exploration of extensions or alternatives to the standard model. 

To address these issues, several alternative cosmological models have been proposed. The alternatives are broadly classified into two classes. The first one considers a dark energy component that evolves with time or an interacting dark energy within the framework of General Relativity (GR). The second category is based on modified or alternative theories of gravity, which attribute the observed accelerated expansion of the Universe to departures from General Relativity  rather than to the presence of an exotic dark energy component.
Among the most notable interacting dark energy models  are those in which a non-gravitational coupling term is introduced to describe a possible interaction between dark matter and dark energy \citep{Dalal,szy,Campo}. These models are based  on the premise that energy exchange between the dark sectors can take place, provided that the total energy-momentum of the combined system remains conserved. In interacting dark energy (IDE) models, the energy transfer between dark matter and dark energy naturally leads to an effective dynamical equation of state for dark energy, which may help alleviate some of the longstanding tensions and challenges associated with the $\Lambda$CDM model. Recently, the Dark Energy Spectroscopic Instruments (DESI) collaboration also reported  hints of dynamical dark energy \citep{Adame} using the Chevallier-Polarski-Linder (CPL) parametrization of the equation of state for the dark energy.

At present,  no fundamental physical theory provides a definitive description of the interaction between dark matter (DM) and dark energy (DE). Consequently, a broad class of phenomenological interacting dark energy (IDE) models has been proposed in the literature. \citet{Amendola}, \citet{Zimdahi}, and \citet{Chimento} were among the first to investigate interacting dark energy scenarios, in which a scalar field is coupled to dark matter through a coupling parameter, and demonstrated that such interactions can alleviate the cosmic coincidence problem without requiring severe fine-tuning.
Various studies explored phenomenological interactions between the dark sectors and placed observational constraints on the coupling parameter using various cosmological probes, showing that a weak interaction remains compatible with current observations \citep{Wang7,He,He2}.  Comprehensive reviews of  IDE models can be found in \citet{Zimdahl,Bolotin}

In this article, we investigate three phenomenological interacting dark energy (IDE) models, each characterized by a distinct functional form of the interaction term, $Q$. These models are denoted as $Q_{1}$, $Q_{2}$, and $Q_{3}$. The $Q_{1}$ and $Q_{2}$ models are linear interaction models, where the interaction term is proportional to the matter energy density, $\rho_{m}$, and the dark energy density, $\rho_{de}$, respectively. The $Q_{3}$ model is based on the relation 
$\rho_{de} \propto \rho_{m}a^{\xi}$, where $\xi$ is a free parameter to be constrained from observations. The special case $\xi = 3$ corresponds to the standard $\Lambda$CDM model.

To constrain the model parameters, we employ several independent low-redshift cosmological probes, including baryon acoustic oscillation (BAO) distance measurements from the
Dark Energy Spectroscopic Instrument (DESI 2024) survey , $H(z)$ measurements obtained from cosmic chronometers, and quasar angular size data. Combining DESI BAO data with cosmic chronometer measurements helps break the degeneracy between the Hubble constant $H_{0}$  and the sound horizon at the drag epoch $r_{d}$, enabling independent constraints on both parameters. The inclusion of QSO angular-size data further improves the constraining power of the combined dataset, yielding tighter and more robust constraints. Using the combined dataset, we investigate whether current observational data support an interaction between dark matter and dark energy and whether such an interaction can alleviate the cosmic coincidence problem.   
 The viability of the proposed phenomenological models is then assessed relative to the standard $\Lambda$CDM model through the use of the Akaike Information Criterion (AIC) and the Bayesian Information Criterion (BIC). The remainder of this paper is structured as follows:
 In Section 2, we introduce the IDE models and discuss their theoretical framework. Section 3 describes the observational datasets used in the analysis. The statistical methods employed for parameter estimation and model comparison are presented in Section 4. The results of the analysis are discussed in Section 5. Finally, Section 6 summarizes the key findings of this study and presents the main conclusions.

\section{Interacting Dark Energy Models}
The observed large scale  homogeneity and isotropy of the Universe imply that  spacetime geometry can be described by the Friedmann-Lema$\hat{\mathrm{i}}$tre-Robertson-Walker (FLRW) metric,
\begin{equation}
ds^{2} = c^{2}dt^{2}-a(t)^{2}[\frac{dr^{2}}{1-K r^{2}} + r^{2}(d\theta^{2}+ \sin^{2}{\theta} d\phi^{2})]
\end{equation}
where $a(t)$ is the scale factor and $K$ characterizes the spatial curvature of the Universe.
In the framework of general relativity (GR), the conservation of energy momentum-tensor, $\nabla_{\mu}T^{\mu\nu}=0$, leads to the continuity equation,
\begin{equation}
\dot{\rho}_{k} + 3H(p_{k}+\rho_{k}) = 0
\end{equation}
where $\rho_{k}$ and $p_{k}$ represent the energy density  and pressure of the $k$th component  (radiation, matter and dark energy), respectively, and $H = \frac{\dot{a}}{a}$ is the Hubble parameter. 
The continuity equation implies that each component of the Universe is conserved independently and there is no energy exchange or interaction among the different components of the Universe.
In interacting dark energy models, the dark matter (DM) and dark energy (DE) components are assumed to exchange energy, while the total energy of the dark sector is conserved.
 This interaction in dark sector is described by an interaction function $Q$ and the continuity equations for DM and DE are modified as,
\begin{equation}
\dot{\rho}_{m} + 3H\rho_{m} = Q
\end{equation}
\begin{equation}
\dot{\rho}_{de} + 3H(p_{de}+\rho_{de}) = -Q
\end{equation}
The sign of $Q$ determines the direction of energy transfer: a positive $Q$ corresponds to energy transfer from dark energy to dark matter, while a negative $Q$ indicates the reverse process.
The continuity equations for radiation and baryonic matter remain unchanged as these components are assumed to be uncoupled from the dark sector interaction.
Since there is currently no fundamental theoretical framework that uniquely determines the form of the interaction term $Q$, a variety of phenomenological models have been proposed based on dimensional arguments, mathematical simplicity, and observational considerations \citep{Campo1,Wang,Feng,Costa}. 
Simple physical considerations and observational constraints suggest that $Q$ should be small.
For example, a large negative $Q$ would cause dark energy to dominate at early times, resulting in accelerated expansion and suppressing the formation of cosmic structures. Hence, viable interacting dark energy models typically require the interaction strength to be sufficiently weak and consistent with observational data.

The continuity equations (3) and (4) suggest that the interaction term $Q$ should be proportional to an energy density multiplied by a quantity with dimensions of inverse time. A natural choice is therefore to consider interaction terms of the form $Q = Q(\rho_{m}H,\rho_{de}H)$. In this work, we investigate two phenomenological  models characterized by  linear interaction terms,
\begin{equation}
Q_{1} = 3\alpha H\rho_{m}
\end{equation} 
and
\begin{equation}
Q_{2} = 3\beta H\rho_{de}
\end{equation}    
where $\alpha$ and $\beta$ are dimensionless  coupling parameters that quantify the strength of interaction between DM and DE. 

We also investigate a third model based on the phenomenological assumption that the dark energy density, $\rho_{de}$, and the dark matter energy density, $\rho_{m}$, are related through \citep{Dalal},
\begin{equation}
\rho_{de} \propto \rho_{m} a^{\xi}
\end{equation}
where $\xi$ is a dimensionless parameter that characterizes the scaling relation between the two dark-sector components.
In a spatially flat FRLW Universe, with the current DE density parameters $\Omega_{de}$ and the current matter energy density parameter $\Omega_{m}$ satisfying the relation: $\Omega_{de}+\Omega_{m}=1$,  one can derive the expression for the interaction term,
\begin{equation}
Q_{3} = -H\rho_{m}(\xi + 3 w)\frac{1-\Omega_{m}}{1-\Omega_{m}+\Omega_{m}(1+z)^{\xi}}
\end{equation}
 where $w \equiv p_{de}/\rho_{de}$ denotes the dark energy equation of state parameter.
The values $\xi = 3$  and $w = -1$ corresponds to the $\Lambda$CDM model.

The expression for the Hubble parameter, $H(z)$, in three phenomenological interacting dark energy models discussed above is obtained by solving the Friedmann equations in a spatially flat FLRW background.

$\bullet$ For $Q_{1} = 3\alpha H\rho_{m}$ :
\begin{equation}
H(z)^{2} = H_{0}^{2}\left[\frac{w\Omega_{m}}{\alpha+w}(1+z)^{3(1-\alpha)} + (1-\frac{w\Omega_{m}}{\alpha+w})(1+z)^{3(1+w)}\right]
\end{equation}

$\bullet$ For $Q_{2} = 3\beta H\rho_{de}$ :
\begin{equation}
H(z)^{2} = H_{0}^{2}\left[(1-\Omega_{m})(1+z)^{3(1+\beta+w)} + \frac{w\Omega_{m}+\beta+\beta(\Omega_{m}-1)(1+z)^{3(\beta+w)}}{(\beta+w)(1+z)^{-3}}\right]
\end{equation}

$\bullet$ For $Q_{3} = -H\rho_{m}(\xi + 3 w)\frac{1-\Omega_{m}}{1-\Omega_{m}+\Omega_{m}(1+z)^{\xi}}$ :
\begin{equation}
H(z)^{2} = H_{0}^{2}\left[(1+z)^{3}[\Omega_{m}+(1-\Omega_{m})(1+z)^{-\xi}]^{-3w/\xi}\right]
\end{equation}
Each  model is specified by four independent parameters, namely ($H_{0},\Omega_{m},w,\alpha$) for $Q_{1}$, ($H_{0},\Omega_{m},w,\beta$) for $Q_{2}$ and ($H_{0},\Omega_{m},w,\xi$) for $Q_{3}$, whose values can be constrained by fitting to cosmological observational data.

\section{Data}

To obtain observational constraints on the model parameters, we use a compilation of cosmic chronometer $H(z)$ measurements together with baryon acoustic oscillation (BAO)  data and quasar angular size observations (QSO).

\subsection{Cosmic Chronometers Data}

Cosmic chronometers provide a novel and  model-independent method for measuring the expansion history of the Universe. The technique relies on determining the differential ages of the oldest passively evolving galaxies (early-type galaxies) located at slightly different redshifts.
In a FRLW Universe, the redshift $z$ is defined as: $1+z = a_{0}/a(t)$, where $a_{0}$ is the present value of the scale factor.  The rate of expansion, represented by the Hubble parameter, as a function of $z$ is,
\begin{equation}
H(z) = \frac{\dot{a}}{a} = -\frac{1}{1+z}\frac{dz}{dt}
\end{equation}
The cosmic chronometer approach directly estimates $H(z)$ by measuring the quantity $dt/dz$, i.e., the age difference of two galaxies separated by a small redshift interval. This age difference is inferred from the amplitude of the  $4000$ \AA~ spectral break ($D_{4000}$) in their absorption spectra, which serves as a reliable indicator of the age of the stellar population.
The selected galaxies are typically very massive early-type systems that evolve passively, exhibiting little or no ongoing star formation. It is assumed that the bulk of their stellar mass formed at high redshifts and that their subsequent evolution has been largely passive. Under these conditions, such galaxies act as reliable cosmic chronometers, enabling direct measurements of the Universe's expansion rate \citep{Morseco}.

We use 32 cosmic chronometers measurements spanning the redshift range $0.07 \leq z \leq 1.965$, compiled by \citet{Morseco}( also by \citet{Barua}). The analysis incorporates the full covariance matrix, $C_{stat+syst}$, to account for both statistical and systematic uncertainties, as recommended by \citet{Morseco20}. The systematic error budget primarily consists of uncertainties arising from the choice of models used for age estimation. These include uncertainties associated with the initial mass function, star formation history, stellar population synthesis models, and stellar metallicity.

\subsection{Baryon Acoustic Oscillations Data}
In the early Universe ($z\geq 1100$), photons and baryons were tightly coupled through Thomson scattering, forming a nearly homogeneous photon–baryon plasma. The competition between radiation pressure and gravitational attraction generated acoustic waves that propagated through this medium. As the Universe expanded and cooled, recombination occurred, leading to the decoupling of photons and baryons. Following this epoch, often referred to as the drag epoch, photons began to free-stream through the Universe and are observed today as the cosmic microwave background radiation (CMBR). The baryon distribution retained the imprint of these acoustic oscillations, producing overdense regions separated by a characteristic co-moving scale, $r_{d}$
, known as the sound horizon at the drag epoch.
 This scale is subsequently imprinted on the large-scale distribution of matter and is observed as a distinct peak in the two-point correlation function of galaxies, commonly referred to as the baryon acoustic oscillation (BAO) feature. By fitting theoretical correlation function templates to the observed galaxy clustering signal, the apparent BAO scale can be measured both perpendicular and parallel to the line of sight.
 The resulting BAO constraints are commonly expressed in terms of the dimensionless distance ratios ($D_{M}/r_{d}$) and ($D_{H}/r_{d}$) or ($D_{v}/r_{d}$), where
\begin{equation}
D_{M}(z) = c  \int_{0}^{z}\frac{dz'}{H(z')}
\end{equation} 
is the co-moving angular diameter distance which determines the transverse BAO scale, and
\begin{equation}
D_{H}(z) = \frac{c}{H(z)}
\end{equation}
is the Hubble distance which characterizes the line of sight BAO scale.
The angle averaged or isotropic BAO scale measurements constrain the combination $D_{v}$, expressed as:
\begin{equation}
D_{v}(z) = \left[z D_{H}(z)D_{M}^{2}(z)\right]^{1/3}
\end{equation}
The co-moving distance $D_{M}(z)$ is related to the physical angular diameter distance, $D_{A}(z)$, as:
\begin{equation}
D_{M}(z) = (1+z)D_{A}(z)
\end{equation}


Recently, the Dark Energy Spectroscopic Instrument (DESI) survey released its first set of baryon acoustic oscillation (BAO) measurements \citep{Adame}, based on spectroscopic observations of more than 18 million unique objects across seven redshift bins spanning $z= 0.1$ to $4.2$.
The DESI Data Release 1 (DR1) provides isotropic and anisotropic BAO measurements derived from several tracers, including luminous red galaxies (LRGs), emission-line galaxies (ELGs), Bright Galaxy Survey (BGS) targets, quasars (QSOs), and the Lyman-$\alpha$ forest.

In this work, we employ  
the latest DESI DR1 BAO measurements and the dataset used in the analysis is summarized in Table 1. The anisotropic BAO measurements are correlated, and their uncertainties are accounted for by using the covariance matrix, $C$, constructed from the measurement uncertainties $\sigma$  and the corresponding correlation coefficients $r$, as reported in Table 1.

\begin{table}

\begin{center}

\begin{tabular}{|c|c|c|c|c|c|}  \hline\hline\
 $z_{eff}$ & Measurements & Value &$\sigma$ &r&Tracer \\ 
\hline & & & & & \\

0.295 & $D_{V}/r_{d}$ & 7.93&0.15 &-& BGS  \\
0.510 & $D_{M}/r_{d}$ & 13.62&0.25 &-0.445 & LRG1 \\
0.510 & $D_{H}/r_{d}$ & 20.98&0.61 &-0.445 & LRG1  \\
0.706 & $D_{M}/r_{d}$ & 16.85&0.32 & -0.420 & LRG2  \\
0.706 & $D_{H}/r_{d}$ & 20.08 & 0.60 & -0.420 & LRG2  \\
0.903 & $D_{M}/r_{d}$ & 21.71&0.28 & -0.389 & LRG3+ELG1  \\
0.903 & $D_{H}/r_{d}$ & 17.88&0.35 & -0.389& LRG3+ELG1 \\
1.317 & $D_{M}/r_{d}$ & 27.79&0.69 & -0.444& ELG2 \ \\
1.317 & $D_{H}/r_{d}$ & 13.82&0.42 & -0.444& ELG2  \\
1.491 & $D_{V}/r_{d}$ & 26.7&0.67 & -& QSO  \\
2.330 & $D_{M}/r_{d}$ & 39.71&0.94 & -0.477& Ly QSO  \\
2.330 & $D_{H}/r_{d}$ & 8.52&0.17 & -0.477&Ly QSO  \\

&&&&&\\
\hline \hline
\end{tabular}

\caption{\small DESI BAO data Release 1\citep{Adame}. The covariance matrix for the anisotropic measurements can be constructed from the uncertainties $\sigma$ and the correlation factor $r$.  }
\end{center}
\end{table}

\subsection{Quasars}
Ultracompact structure in radio quasars have recently emerged as reliable cosmological probes and can be employed as standard rulers for constraining cosmological models \citep{Vish,Lima,Chen3}.
Observations indicate that the linear size of compact structures in intermediate-luminosity quasars exhibits negligible dependence on both redshift and intrinsic luminosity. Consequently, a suitably selected population of these objects can be treated as standard rulers characterized by a nearly constant intrinsic length scale, $l_{m}$ \citep{Cao17}. For an object at redshift $z$, the intrinsic linear size $l_{m}$ 
is related to its observed angular size $\theta(z)$ and angular diameter distance 
$D_{A}(z)$ through
\begin{equation}
\theta(z) = \frac{l_{m}}{D_{A}(z)}
\end{equation}

\citet{Cao17a} compiled a subsample of 120 intermediate-luminosity quasars, spanning the redshift range of $0.46 < z < 2.8$, from a catalog of 613 ultra compact radio sources imaged  by the Very-Long Baseline Interferometry (VLBI) all-sky survey at $2.29$ GHz. The physical linear size of the quasars in the subsample show negligible dependence on redshift and intrinsic luminosity. Using a model independent calibration technique the authors constrained the linear intrinsic size to $l_{m} =11.02 \pm 0.25$pc, corresponding to the typical radius at which active galactic nuclei  (AGN) jets become opaque at the observed frequency $\nu \approx 2$ GHz. Employing this subsample, they obtained stringent constraints on cosmological parameters. For our analysis, we use the angular size measurement $\theta(z)$ in milliseconds, and the corresponding redshift values listed in Table 1 of \citet{Cao17a}.

\section{Method}

To estimate the best-fitting model parameters we adopt  Bayesian Statistics, that treats the unknown parameters as random variables. Bayesian inference combines prior knowledge about the parameters (the prior distribution) with the observed data to obtain the posterior distribution. The posterior distribution represents the probability of observing the data given a specific set of model parameters. It quantifies the updated uncertainty in the model parameters after incorporating the observational data. It provides insight into the precision with which the parameters are constrained and reveals potential correlations and dependencies among them.

Unlike the standard least squares methods, the Bayesian approach explores the entire parameter space to build the posterior distribution. The likelihood function for a model with parameter vector, $\textbf{p}$, and given data, $D$, is defined as:
\begin{equation}
\textit{L}(D | \textbf{p}) = e^{-\chi^{2}(\textbf{p})/2}
\end{equation}
The form of the $\chi^{2}$ function depends on the dataset under consideration. For the datasets used in our analysis, the corresponding $\chi^{2}$ functions are defined as follows:

$\bullet$ For $H(z)$ data points:
\begin{equation}
\chi^{2}_{CC}(\textbf{p}) = [\mathbf{\bigtriangleup H}] ^{T}\cdot C_{cc}^{-1}\cdot [\mathbf{\bigtriangleup H}] 
\end{equation} 
where $\mathbf{\bigtriangleup H}_{i} \equiv (H_{th}(\textbf{p};z_{i})-H_{ob}(z_{i}))$ is the residual vector,
 $H_{th}(\textbf{p})$ denotes the Hubble parameter expression for the specific model, $H_{ob}$ is the observed value of Hubble parameter and $C_{cc}$ is  the $32$x$32$ covariance matrix for the $H(z)$ measurements.

$\bullet$ For the uncorrelated BAO data points:
\begin{equation}
\chi^{2}_{BAO} (\textbf{p}) = \sum_{i=1}^{N} \frac{[D_{th}(\textbf{p};z_{i})-D_{ob}(r_{d};z_{i}]^{2}}{\sigma_{i}^{2}} 
\end{equation}
here $N$ is the number of data points, $D_{th}$ and $D_{ob}$ are, respectively, the theoretical and observed value of the distance ratios as listed in Table 1, and $\sigma_{i}$ is the uncertainty corresponding to $D_{ob}(z_{i})$.  The BAO measurements  involve the sound horizon $r_{d}$, at the drag epoch, which  we treat as a free parameter (instead of using the standard fit function as proposed in \citet{Eisenstein},
to avoid potential biases) and constrain it jointly with the other model parameters.

For the correlated BAO data:
\begin{equation}
\chi^{2}_{BAO}(\textbf{p}) = [D_{th}(\textbf{p};z_{i})-D_{ob}(r_{d};z_{i})] ^{T}\cdot C_{BAO}^{-1}\cdot [D_{th}(\textbf{p};z_{i})-D_{ob}(r_{d};z_{i})] 
\end{equation} 

$\bullet$ For the QSO data:
\begin{equation}
\chi^{2}_{QSO} (\textbf{p}) = \sum_{i=1}^{120} \frac{\left[\theta_{th}(\textbf{p};z_{i})-\theta_{ob}(z_{i})\right]^{2}}{\sigma_{i}^{2}+(0.1\theta_{ob}(z_{i}))^{2}} 
\end{equation}
where $\theta_{th}(z_{i})$ and $\theta_{ob}(z_{i})$ denote the model-predicted and the measured angular sizes at redshift $z_{i}$, respectively, while $\sigma_{i}$ represents the corresponding measurement uncertainty. Following \citep{Cao17a}, an additional systematic uncertainty of $0.1\theta_{ob}(z_{i})$ is added to $\sigma_{i}$ in quadrature to account for both observational (statistical) errors and the intrinsic spread in linear sizes (systematic errors).
 
The Total likelihood function for the joint analysis of the $H(z)$, BAO and QSO datasets is given by
\begin{equation}
\mathcal{L}_{Tot} = \mathcal{L}_{cc} \mathcal{L}_{BAO} \mathcal{L}_{QSO}
\end{equation}
The posterior probability density function (PDF) for the parameter vector $\textbf{p}$ for the combined data, $D$, is 
\begin{equation}
\textit{L}(\textbf{p}|D) \propto \mathcal{L}_{Tot}  \pi(\textbf{p})
\end{equation}
where $\pi(\textbf{p})$ denotes the prior distribution of the model parameters. The one-dimensional (1D) marginalized posterior PDF of a given parameter is obtained by integrating the joint posterior over all remaining parameters.

We employ the Python package EMCEE \citep{emcee}, an affine-invariant ensemble Markov Chain Monte Carlo (MCMC) sampler, to generate samples from the posterior distribution of the model parameters. The priors adopted for the analysis are listed in Table 2. From the converged MCMC chains, we obtain the  parameter constraints by using the sample means as the central values and deriving the corresponding $1\sigma$ ($68\%$) credible intervals from the marginalized posterior distributions. The one-dimensional marginalized posterior distributions and the two-dimensional (2D) joint posterior confidence contours are generated using the GETDIST package \citep{getdist}.

The goodness of fit of the models is compared using two widely adopted model selection criteria: the Akaike Information Criterion (AIC) \citep{AIC} and the Bayesian Information Criterion (BIC) \citep{BIC}.
The AIC and BIC values for a model with $k$ number of parameters fitted with $N$ data point are defined as,
\begin{equation}
\mathrm{AIC} = \chi_{min}^{2} + 2k
\end{equation}
for $N >> k$, and
\begin{equation}
\mathrm{BIC} = \chi_{min}^{2} + k \mathrm{ln}N
\end{equation}
where $\chi_{min}^{2}$ is the minimum value of $\chi^{2}$ in the given model.


\begin{table}[htb] 
\caption{The prior distributions of parameters used in the analysis.} 
\begin{center}

\begin{tabular}{|c|c| }
\hline
\textbf{Parameter}  & \textbf{Priors} \\
\hline
&\\
$H_{0}$ & Uniform ($50 , 100)$ )\\
&\\
$\Omega_{m}$ & Uniform ($0.01 , 0.99)$ )\\
&\\
$w$ & Uniform ($-2 , 0)$ )\\
&\\
 $\alpha$ & Uniform ($-2 , 2)$ )\\
&\\
$\beta$ & Uniform ($-2 , 2)$ )\\
&\\
$\xi$ & Uniform ($0 , 6)$ )\\
&\\
$r_{d}$ & Uniform ($100, 200)$ )\\
&\\
\hline
\end{tabular} \\
\end{center}

\end{table}

\section{Results}

Using the combined dataset of $32$ $H(z)$ measurements, $12$ BAO distance measurements and $120$ angular size measurements of quasars, we obtain the constraints on the model parameters and  $r_{d}$, the sound horizon at the drag epoch.
The sample means and two sided uncertainties ($16$th and $84$th percentile values) of the marginalized parameters for two different data combinations are presented in Table 3. The $1D$ marginalized posterior probability distribution function and the $2D$ confidence contours for the parameters are shown in Fig. $1-4$.

\begin{table}[htb] 
\begin{center}

\caption{Constraints on cosmological parameters for the IDE models and the Flat $\Lambda$CDM model.  Hubble parameter $H_{0}$ is expressed in units of Km $\mathrm{s}^{-1} \mathrm{Mpc}^{-1}$  and $r_{d}$ is in Mpc.}

$Q_{1}$ Model

\begin{tabular}{cccccc}

&&&&&\\
Data &$H_{0}$  & $\Omega_{m}$ & $w$ & $\alpha$&$r_{d}$\\
\hline
&&&&&\\
BAO+CC&$68.500^{ +3.300}_{ -3.086}$ & $0.230^{ +0.122}_{ -0.118}$ & $-0.788^{ +0.211}_{ -0.395}$ & $-0.091^{+0.148}_{-0.242}$&$144.981^{+4.991}_{-4.815}$\\
&&&&&\\
BAO+CC+QSO&$67.520^{ +2.511}_{ -1.814}$ & $0.225^{ +0.134}_{ -0.126}$ & $-0.779^{ +0.217}_{ -0.408}$ & $-0.0991^{+0.159}_{-0.274}$&$146.422^{+1.895}_{-1.808}$\\
&&&&&\\
\hline
\end{tabular} \\

$Q_{2}$ Model

\begin{tabular}{cccccc}

&&&&&\\
Data &$H_{0}$  & $\Omega_{m}$ & $w$ & $\beta$&$r_{d}$\\
\hline
&&&&&\\
BAO+CC&$69.650^{ +3.034}_{ -3.114}$ & $0.389^{ +0.167}_{ -0.229}$ & $-1.148^{ +0.325}_{ -0.413}$ & $0.146^{+0.411}_{-0.312}$&$145.071^{+5.202}_{-4.907}$\\
&&&&&\\
BAO+CC+QSO&$68.509^{ +2.520}_{ -2.011}$ & $0.416^{ +0.165}_{ -0.236}$ & $-1.182^{ +0.356}_{ -0.445}$ & $0.189^{+0.452}_{-0.333}$&$146.541^{+1.884}_{-1.765}$\\
&&&&&\\
\hline
\end{tabular} \\

$Q_{3}$ Model

\begin{tabular}{cccccc}

&&&&&\\
Data &$H_{0}$  & $\Omega_{m}$ & $w$ & $\xi$&$r_{d}$\\
\hline
&&&&&\\
BAO+CC&$69.096^{ +3.214}_{ -3.009}$ & $0.333^{ +0.198}_{ -0.194}$ & $-1.051^{ +0.349}_{ -0.566}$ & $2.910^{+0.879}_{-0.624}$&$145.268^{+5.160}_{-4.650}$\\
&&&&&\\
BAO+CC+QSO&$68.317^{ +2.286}_{ -2.144}$ & $0.320^{ +0.191}_{ -0.194}$ & $-1.021^{ +0.351}_{ -0.518}$ & $2.975^{+0.857}_{-0.653}$&$146.588^{+1.803}_{-1.804}$\\
&&&&&\\
\hline
\end{tabular} \\

Flat $\Lambda$CDM Model

\begin{tabular}{cccccc}
&&&&&\\
Data &$H_{0}$  & $\Omega_{m}$&$r_{d}$&&\\
\hline
&&&&&\\
BAO+CC&$69.518^{ +2.456}_{ -2.472}$ & $0.301^{ +0.015}_{ -0.014}$ &$145.350^{+5.420}_{-4.749}$&&
\\
&&&&&\\
BAO+CC+QSO & $68.856^{ +1.050}_{ -1.040}$ & $0.303^{ +0.015}_{ -0.014}$ &  $146.604^{+1.852}_{-1.864}$ &&\\
&&&&&\\
\hline
\end{tabular} \\

\end{center}
\end{table}

\begin{table}[htb] 
\caption{The $\chi^{2}_{min}$, AIC and BIC values for the three IDE models and Flat $\Lambda$CDM model.} 
\begin{center}

\begin{tabular}{cccccc }
\hline
\textbf{Model}  & $\chi^{2}_{min}$ & AIC & BIC & $\bigtriangleup$AIC & $\bigtriangleup$ BIC\\
\hline
&&&&&\\
$Q_{1}$ & $636.813$ &$646.813$& $662.312$ &$2.351$  & $8.551$ \\
&\\
$Q_{2}$ & $638.376$ &$648.376$& $663.875$  & $3.914$ & $10.114$\\
&\\
$Q_{3}$ & $638.786$ &  $648.786$ &$664.285$ &  $4.324$ & $10.524$\\
&\\
Flat $\Lambda$CDM & $638.462$ &$644.462$& $653.761$ &--&-- \\
&\\

\hline
\end{tabular} \\
\end{center}

\end{table}

\begin{figure}[p]
\centering

  \includegraphics[width=1.0\linewidth]{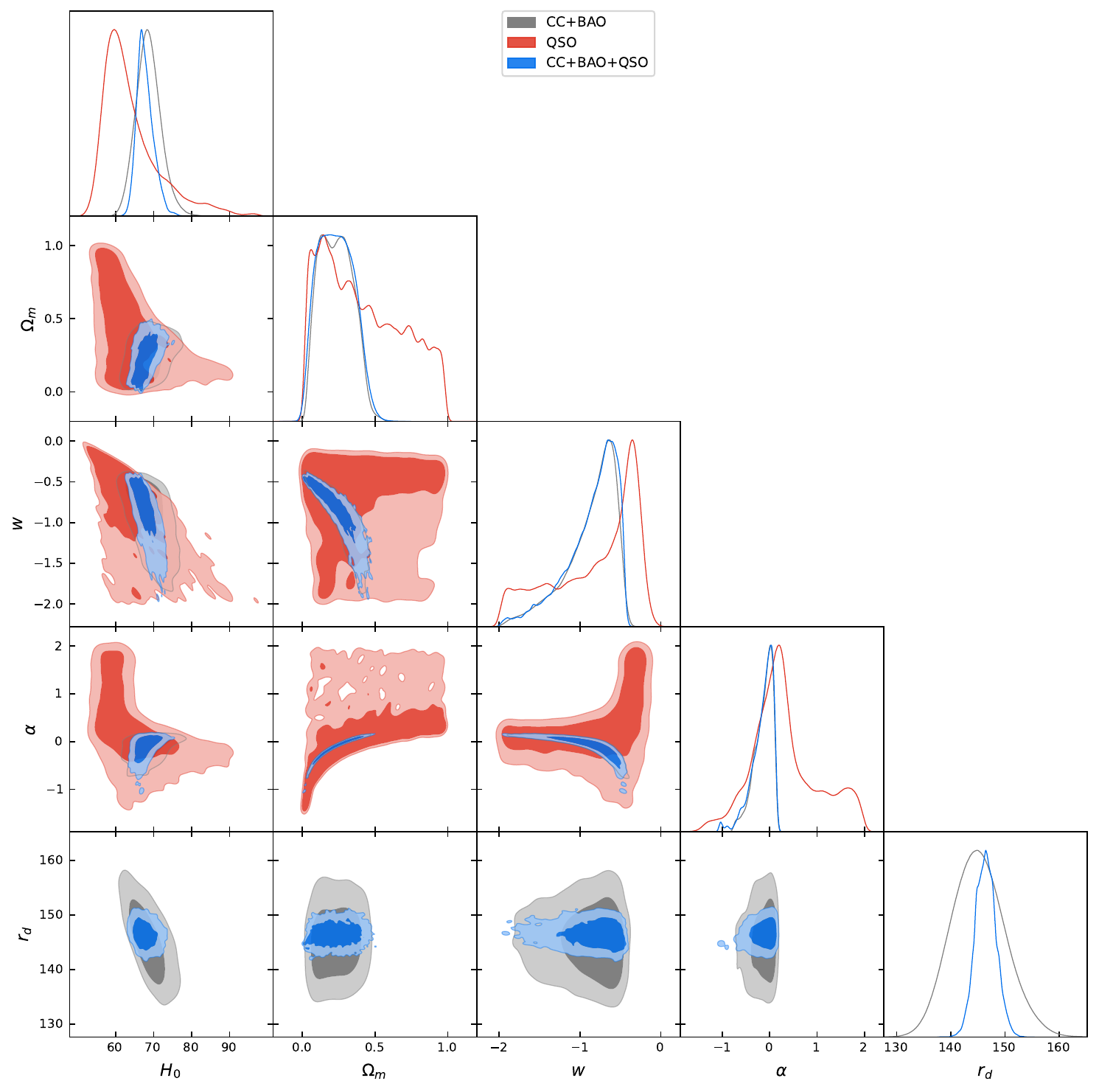}
 \caption{ \small{ Constraint results for the parameters of $Q_{1}$ IDE model using $H(z)$, BAO and QSO measurements.}}
  
\end{figure}

\begin{figure}[p]
\centering

  \includegraphics[width=1.0\linewidth]{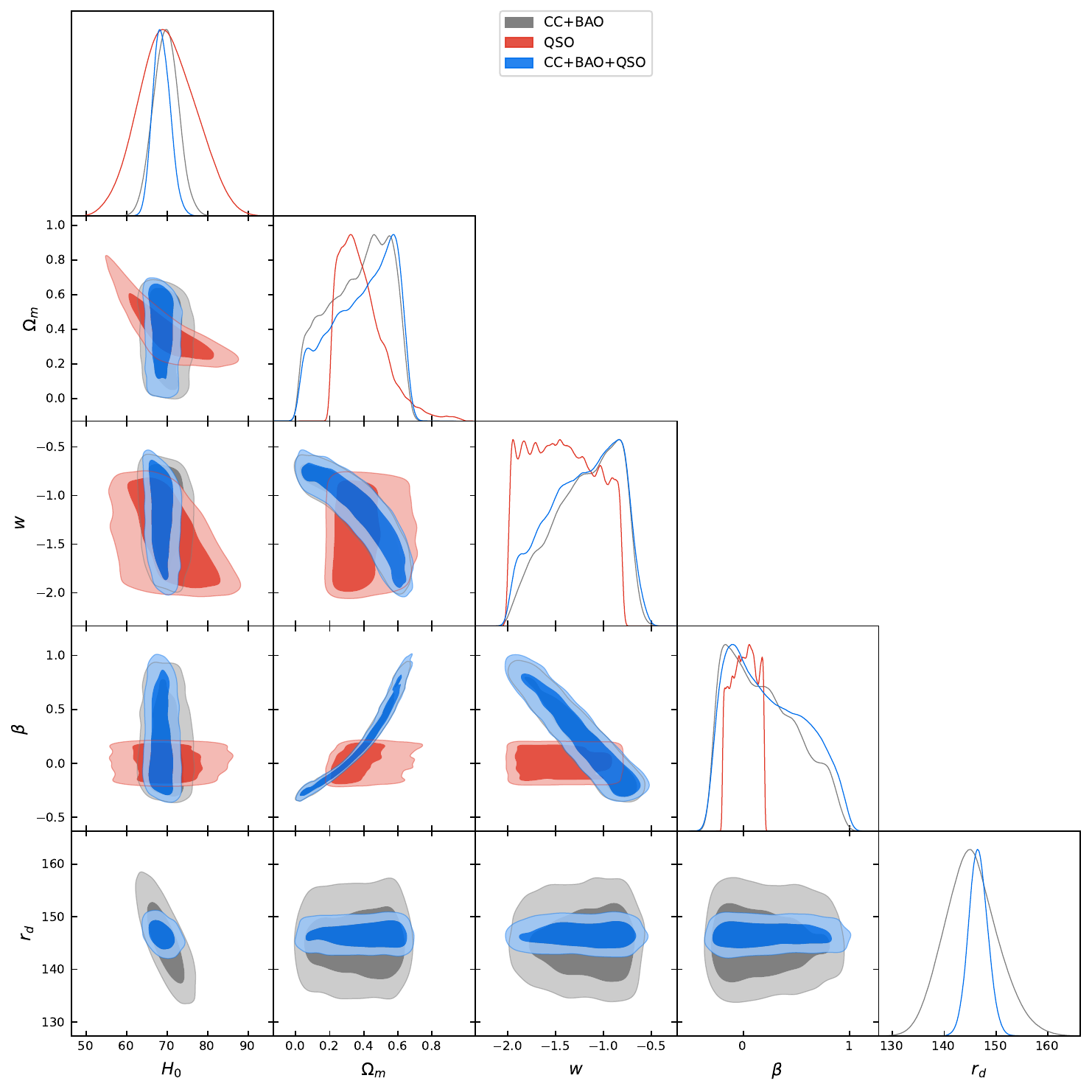}
 \caption{ \small{ Constraint results for the parameters of $Q_{2} $IDE model using $H(z)$, BAO and QSO measurements.}}
  
\end{figure}

\begin{figure}[p]
\centering

  \includegraphics[width=1.0\linewidth]{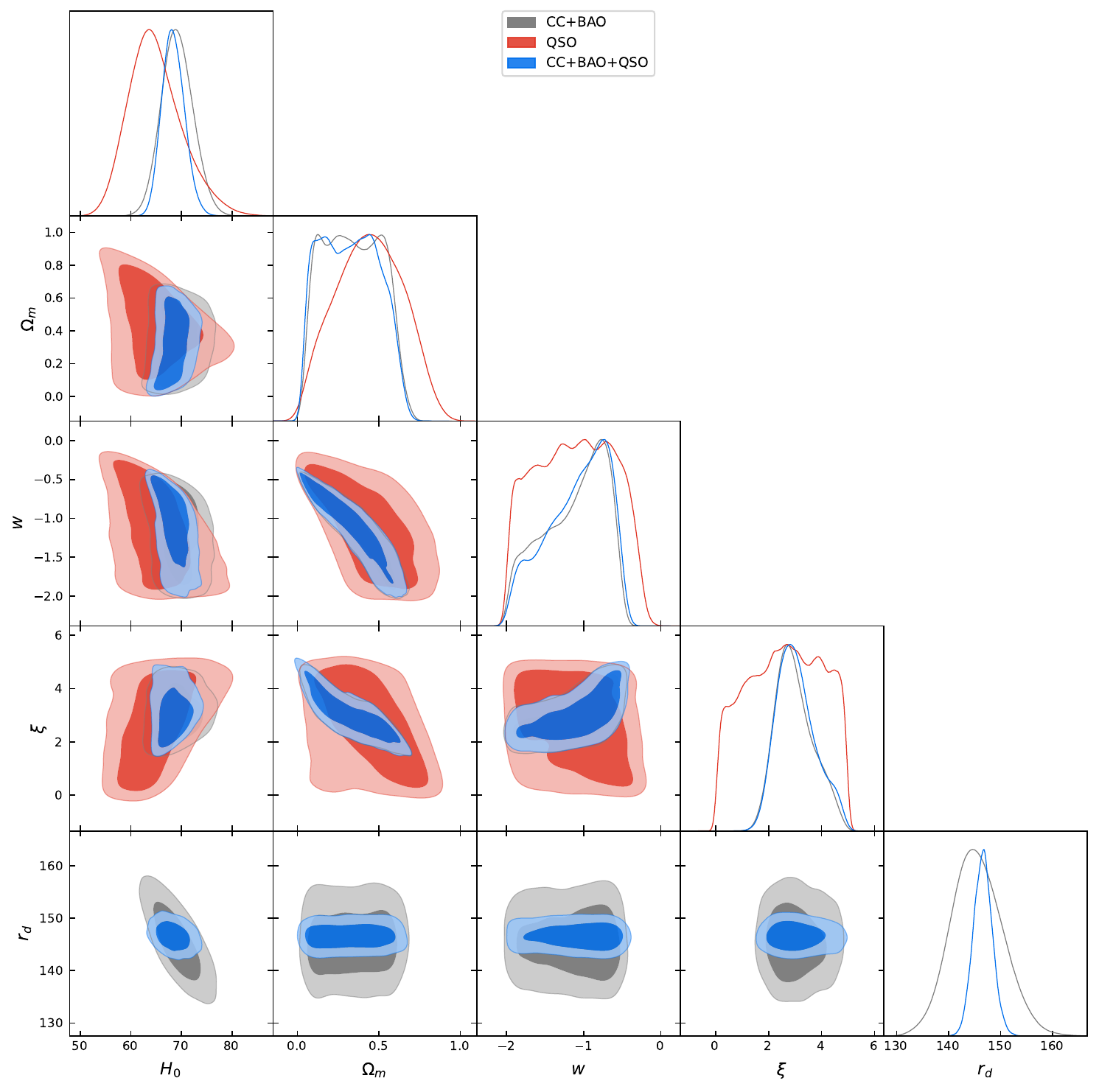}
 \caption{ \small{ Constraint results for the parameters of $Q_{3}$ IDE model using $H(z)$, BAO and QSO measurements.}}
  
\end{figure}

\begin{figure}[p]
\centering

  \includegraphics[width=1.0\linewidth]{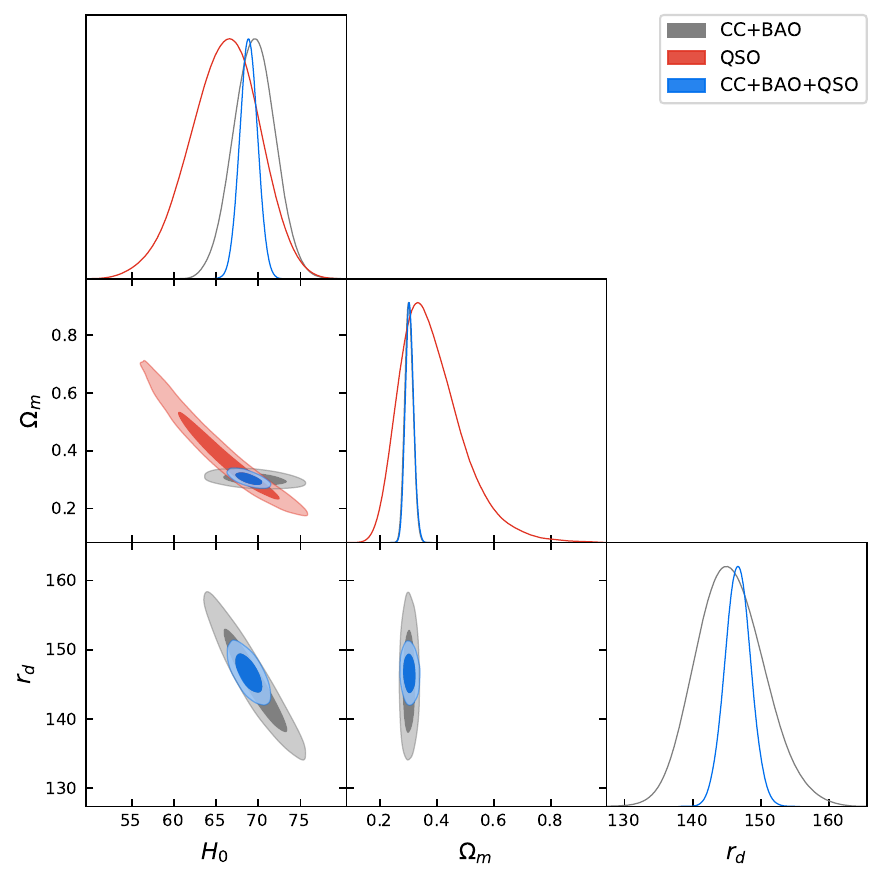}
 \caption{ \small{ Constraint results for the parameters of $\Lambda$CDM model using $H(z)$, BAO and QSO measurements.}}
  
\end{figure}

$\bullet$ For the $Q_{1}$ interaction model, defined by $Q_{1} = 3\alpha H \rho_{dm}$, the free parameters constrained in the analysis are $H_{0},\Omega_{m},w,\alpha$ and $r_{d}$. Using the combined CC, BAO, and QSO data set, the interaction parameter is constrained to $\alpha = -0.099^{ +0.159}_{ -0.274}$ at the $1\sigma$ confidence level, suggesting a slight preference for energy transfer from dark matter to dark energy. However, the coincidence problem remains unresolved.
 The dark energy equation-of-state parameter is constrained to $w = -0.779^{ +0.217}_{ -0.408}$, which is marginally greater than $-1$, suggesting a mild preference for a quintessence-like dark energy scenario.  Nevertheless, the $1\sigma$ confidence intervals for both parameters encompass their respective $\Lambda$CDM values, $\alpha =0$ and $w = -1$, indicating that the results remain consistent with the standard $\Lambda$CDM model. Therefore, the observational datasets employed in this analysis do not provide statistically significant evidence to distinguish the $Q_{1}$interacting dark energy model from the standard $\Lambda$CDM cosmology.
Furthermore, the analysis yields $H_{0} = 67.52^{+2.511}_{-1.814}$Km $\mathrm{s}^{-1} \mathrm{Mpc}^{-1}$ and  $\Omega_{m} = 0.225^{ +0.134}_{ -0.141126}$. 
Both $H_{0}$ and $\Omega_{m}$ values are  consistent within the $1\sigma$ confidence interval with the corresponding values obtained for the flat $\Lambda$CDM model  by \citet{Aghanim}, $H_{0} = 67.4\pm 0.5$ Km$\mathrm{s}^{-1} \mathrm{Mpc}^{-1}$ and $\Omega_{m} = 0.3153\pm +0.0073$,  using the temperature and polarization anisotropy measurements of the cosmic microwave radiation (CMB) from the Planck satellite.

The SH0ES collaboration \citep{Riess22} measured $H_{0} = 73.04 \pm 1.04$ Km $\mathrm{s}^{-1} \mathrm{Mpc}^{-1}$ using  local Type 1a supernovae (SNe1a), calibrated through the cosmic distance ladder based on the Hubble-Lema$\hat{\mathrm{i}}$tre law, independent of any cosmological model.
The discrepancy between the Planck CMB constraint (within the framework of the flat $\Lambda$CDM
 model) and the SH0ES measurement corresponds to a tension of $4.89\sigma$. In contrast, the $H_{0}$ value inferred from the $Q_{1}$ model exhibits a reduced tension of $2.03\sigma$ \footnote{the upper uncertainty of the inferred value is adopted for the calculation.} with respect to the SH0ES measurement, primarily due to the larger uncertainties associated with the model's constraint.


$\bullet$  The $Q_{2}$ interaction model, defined by $Q_{2} = 3\beta H \rho_{de}$, contains four free cosmological parameters ( $H_{0},\Omega_{m},w,\beta$ ) in addition to the sound horizon at the drag epoch $r_{d}$. The analysis constraints  the interaction parameter to a value $\beta = 0.189^{ +0.452}_{ -0.333}$ ($68\%$ confidence level) suggesting a mild preference for energy  transfer from dark energy to dark matter. The dark energy equation of state parameter is constrained to $-1.182^{ +0.356}_{ -0.445}$. Although the positive central value of $\beta$ implies that the coincidence problem may be slightly alleviated, both $\beta = 0$ and $w = -1$ lie within the $1\sigma$ confidence interval. 
 The  value of $H_{0} = 68.509^{+2.52}_{-2.011}$Km $\mathrm{s}^{-1} \mathrm{Mpc}^{-1}$ obtained in the $Q_{2}$ model is consistent  within $1\sigma$ confidence interval with the value inferred from Planck CMB measurements for the flat $\Lambda$CDM model. Similarly,
the value of  $\Omega_{m} = 0.416^{ +0.165}_{ -0.36}$, constrained in the $Q_{2}$ model, is   consistent  with the corresponding value obtained for the flat $\Lambda$CDM model  by \citet{Aghanim}, given its relatively large uncertainty. Furthermore, the $H_{0}$
value predicted by the $Q_{2}$  model differs from the SH0ES measurement by approximately $1.67 \sigma$.  This relatively small discrepancy is primarily attributable to the larger uncertainty associated with the $Q_{2}$ estimate. Therefore, the current dataset does not provide statistically significant evidence to distinguish the $Q_{2}$ interacting dark energy model from the standard $\Lambda$CDM cosmology.

$\bullet$ For the $Q_{3}$ model, defined by the interaction term 
$Q_{3} = -H\rho_{dm}(\xi + 3 w)\frac{1-\Omega_{dm}}{1-\Omega_{dm}+\Omega_{dm}(1+z)^{\xi}}$, the constraint results for the free parameters are : $H_{0} = 68.317^{ +2.286}_{ -2.144},\Omega_{m} = 0.320^{ +0.191}_{ -0.94}, w = -1.021^{ +0.351}_{ -0.518} ,\xi = 2.975^{ +0.857}_{ -0.653}$ and $r_{d} = 146.588^{ +1.803}_{ -1.804}$.
From the central values of $w$ and $\xi$,  the mean value of the effective coupling parameter, $\xi + 3w = -0.088$, indicates that the dark energy is converted to the dark matter, implying a slight alleviation of the coincidence problem. However, both $w = -1$ and $\xi = 3$, corresponding to the standard $\Lambda$CDM scenario, lie within the $1\sigma$ confidence interval. Therefore, the dataset does not provide statistically significant evidence to distinguish the $Q_{3}$ interacting dark energy model from the standard $\Lambda$CDM model. The  values of $H_{0}$ and $\Omega_{m}$ obtained, for $Q_{3}$ model, are also  
consistent, within the $1\sigma$ confidence interval, with the corresponding values obtained by \citet{Aghanim} for the flat $\Lambda$CDM model. The estimated $H_{0}$ for the $Q_{3}$ model deviates from the SH0ES measurements by approximately $1.92\sigma$.

$\bullet$ For comparison, we also constrain the parameters of the flat  $\Lambda$CDM model. The resulting constraints on $H_{0}$, $\Omega_{m}$ and $r_{d}$ are presented at the bottom of Table 3. We find that the values of $H_{0}$ and $\Omega_{m}$ obtained for all three interacting dark energy models are consistent, within the $1\sigma$
confidence interval, with those inferred for the flat $\Lambda$CDM  model. As expected, the best-fit value of the sound horizon at the drag epoch, $r_{d}$, remains nearly unchanged across all models. Furthermore, the inferred values of $r_{d}$ are consistent within $1\sigma$
with the CMB constraint of $r_{d} = 147.09 \pm 0.26$ Mpc reported by \citet{Aghanim}.

$\bullet$ We observe that incorporating the QSO data into the CC+BAO dataset tightens the constraints on $H_{0}$ and $r_{d}$ for all three IDE models. However, no significant improvement is observed in the constraints on the parameters $w$ and $\xi$ of the $Q_{3}$
 model. The central values of the interaction parameters $\alpha$ and $\beta$, which characterize the strength of the interaction between dark matter and dark energy , shift slightly toward higher values, while the widths of their respective $1\sigma$ confidence intervals remain nearly unchanged.   
 
$\bullet$ We repeat the analysis using different flat prior ranges for $r_{d}$ and find that the constraints on $H_{0}$ and the IDE parameters remain unchanged. This indicates that combining the QSO data with the CC+BAO dataset effectively breaks the degeneracy between $H_{0}$ and $r_{d}$.

$\bullet$ To identify the most statistically robust model, we compute the minimum chi-square, $\chi^{2}_{min}$, along  with the corresponding AIC and BIC values, which are presented in Table 4.  The $Q_{1}$ model yields the lowest $\chi^{2}_{min}$ , whereas the corresponding values for the $Q_{2}$ and $Q_{3}$ models are marginally higher than that of the flat $\Lambda$CDM model. Nevertheless, the flat $\Lambda$CDM model exhibits the lowest AIC and BIC values,  owing to its smaller number of free parameters. Thus, despite achieving a better or comparable fit to the data, the IDE models are penalized by both information criteria for their additional parameters. Moreover, the positive $\bigtriangleup$AIC and $\bigtriangleup$BIC values for all three IDE models, relative to the flat $\Lambda$CDM model, indicate a preference for the latter. Therefore, based on the AIC and BIC criteria, the dataset considered in this analysis favours the flat $\Lambda$CDM model over the three IDE models, despite the marginally better goodness of fit achieved by the $Q_{1}$ model.

\section{Conclusions and Discussion} 
In this work, we investigated the coincidence problem by considering three phenomenological interacting dark energy (IDE) models. We constrained the model parameters using a combination of cosmic chronometer $H(z)$ data and two standard rulers, namely baryon acoustic oscillation (BAO) and quasar (QSO) angular size observations. The parameter estimation was performed using the Markov Chain Monte Carlo (MCMC) technique to sample the likelihood function, from which we obtained the marginalized posterior means and the corresponding two-sided $1\sigma$  uncertainties. To assess the constraining power of QSO data, we compare the results obtained from two data combinations: CC+BAO and CC+BAO+QSO. The results are presented in Table 3 and Figures 1–4. We find that incorporating QSO data into the CC+BAO dataset helps break the degeneracy between $H_{0}$ and $r_{d}$, yielding tighter and more robust constraints on both parameters across all three IDE models. However, no significant improvement is observed in the constraints on the remaining model parameters. These findings highlight the potential of QSO data to serve as a powerful complementary probe to BAO data, particularly in constraining $H_{0}$ and $r_{d}$ independently.

The marginalized parameter estimates indicate a positive interaction parameter $(Q > 0)$ for 
$Q_{2}$ and $Q_{3}$ models, implying an energy transfer from dark energy to dark matter. Such an interaction can help alleviate the coincidence problem. The $Q_{1}$ model with a negative interaction parameter do not alleviate the coincidence problem.  
  Nevertheless, the standard $\Lambda$CDM model, corresponding to the limiting case $Q \rightarrow 0$, remains consistent with the observational data, as it lies within the $1\sigma$ confidence region of the inferred interaction parameters.

The constraint results for the Hubble constant, $H_{0}$, and the matter density parameter, 
$\Omega_{m}$, are consistent with the high-redshift measurements from the cosmic microwave background (CMB) radiation within the framework of the spatially flat $\Lambda$CDM model. Furthermore, The inferred values of $H_{0}$ differ from the low-redshift, model-independent SH0ES measurement by $2.03\sigma$, $1.67\sigma$, and $1.92\sigma$ for the $Q_1$, $Q_2$, and $Q_3$ models, respectively.  The reduced $H_{0}$ tension exhibited by these IDE models is attributed primarily to the larger uncertainties in the inferred parameter constraints.
Model comparison based on the Akaike Information Criterion (AIC) and the Bayesian Information Criterion (BIC) indicates that the dataset considered in this analysis favours the standard flat 
$\Lambda$CDM model over the three IDE models. Although the IDE models achieved a better or comparable fit to the data, the IDE models are penalized by both information criteria for their additional parameters.

Recently, \citet{Diao} constrained the $\xi$IDE model ($Q_{3}$ model) using strong gravitational lensing (SGL), Type Ia supernovae (SNe Ia), and $H(z)$ data, obtaining $w = -1.24 \pm 0.61$ and $\xi = 3.8 \pm 3.9$. Similarly, \citet{Nong} combined gamma-ray burst (GRB) observations with the Pantheon+ SNe Ia sample and reported $w = -1.55^{+1.0}_{-0.57}$ and $\xi = 1.78^{+0.31}_{-1.10}$.
In comparison, our analysis yields tighter constraints on the $Q_{3}$
 model using the combined CC+BAO+QSO dataset, with $w = -1.021^{ +0.351}_{ -0.518}$  and $\xi = 2.975^{+0.857}_{-0.653}$. 
 Furthermore, \citet{Li} investigated interacting dark energy models using DESI BAO, SNe Ia, and cosmic microwave background (CMB) data, taking the $Q_{1}$ and $Q_{2}$ interaction scenarios with the equation-of-state parameter fixed at $ w = -1$. They obtained constraints on the interaction parameters of: $-0.0003\pm 0.0011$ and $0.197\pm 0.071$ for the $Q_{1}$ and $Q_{2}$ models, respectively.  Similarly, \citet{Giare} analysed the $Q_{2}$ model  using a combination of DESI BAO, Planck 2018 CMB, and CC data, fixing $w = -1$,  and obtained $\beta = 0.32_{-0.14}^{+0.18}$. These results are broadly consistent with our findings and suggest that current observational data favour a weak interaction between dark matter and dark energy.

We conclude that the current dataset support a  mild interaction between the dark sectors and the $Q_{2}$ and $Q_{3}$ phenomenological IDE models show potential for alleviating the coincidence problem. However, the relatively large $1\sigma$ uncertainties associated with the parameters $\alpha$, $\beta$, $\xi$, and $w$ indicate that the current dataset does not provide statistically significant evidence to distinguish these models from the standard $\Lambda$CDM scenario. Consequently, larger and more precise observational datasets will be required for future works to obtain tighter constraints on the model parameters and to draw more definitive conclusions regarding the viability of interacting dark energy models.

\section*{Declarations}

\textbf{Data Availability Statement:} The datasets used in the work are available in public domain. \linebreak
\textbf{Funding and/or Conflicts of interests/Competing interests} This work is done independently and no funds, grants, or other support was received.  There are no conflicts of interests that are relevant to the content of this article.

\section*{Acknowledgements} 
 The authors sincerely acknowledge the anonymous referee's constructive comments and suggestions, which have contributed to improving the quality of the manuscript.

\label{lastpage}

\end{document}